# Synchrotron radiation based beam diagnostics at the Fermilab Tevatron[1]

R. Thurman-Keup,[2] H. W. K. Cheung, A. Hahn, P. Hurh, E. Lorman,[3] C. Lundberg, T. Meyer, D. Miller,[4] S. Pordes and A. Valishev

*Fermi National Accelerator Laboratory,*
*P.O. Box 500  Batavia, IL  60510, USA*

*E-mail*: keup@fnal.gov

ABSTRACT: Synchrotron radiation has been used for many years as a beam diagnostic at electron accelerators.  It is not normally associated with proton accelerators as the intensity of the radiation is too weak to make detection practical.  However, if one utilizes the radiation originating near the edge of a bending magnet, or from a short magnet, the rapidly changing magnetic field serves to enhance the wavelengths shorter than the cutoff wavelength, which for more recent high energy proton accelerators such as Fermilab's Tevatron, tends to be visible light.  This paper discusses the implementation at the Tevatron of two devices.  A transverse beam profile monitor images the synchrotron radiation coming from the proton and antiproton beams separately and provides profile data for each bunch.  A second monitor measures the low-level intensity of beam in the abort gaps which poses a danger to both the accelerator's superconducting magnets and the silicon detectors of the high energy physics experiments. Comparisons of measurements from the profile monitor to measurements from the flying wire profile systems are presented as are a number of examples of the application of the profile and abort gap intensity measurements to the modelling of Tevatron beam dynamics.

---

[1] Work supported by Fermi Research Alliance, LLC under Contract No. DE-AC02-07CH11359 with the United States Department of Energy.
[2] Corresponding author.
[3] No longer at Fermilab.
[4] Retired.

# Contents

## 1. Introduction

Synchrotron radiation is a term given to the radiation that is emitted by a charged particle that undergoes acceleration through the bending magnets of a circular accelerator. The concept was first considered by A. Liénard [1] and E. Wiechert [2] around 1900 with subsequent calculations by Schott [3] in 1912. Further theoretical work was performed in the 1940's by Iwanenko, Pomeranchuk, and Schwinger [4-7] and the radiation was first directly observed in 1947 at the General Electric 70 MeV electron synchrotron [8]. Since then, various aspects of synchrotron radiation have been used as beam diagnostics in electron accelerators.

Synchrotron radiation from proton accelerators is nowhere near as copious. From Jackson [9], synchrotron radiation has the following frequency spectrum

$$\frac{dI}{d\omega} = 2\sqrt{3}\frac{e^2}{c}\gamma\frac{\omega}{\omega_c}\int_{2\omega/\omega_c}^{\infty} K_{5/3}(x)\,dx \qquad (1.1)$$

where $\omega_c = 3\gamma^3(c/\rho)$ is the cutoff frequency, above which there is virtually no emission (see figure 1).

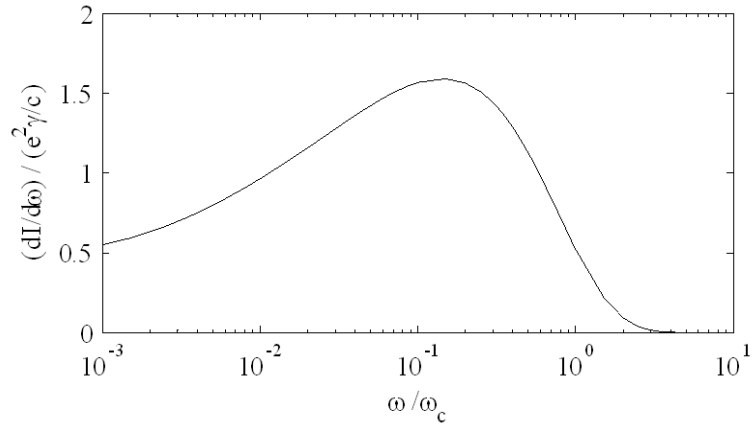

**Figure 1.** Spectrum of synchrotron radiation from a continuous magnetic field.

Since the cutoff frequency depends on $\gamma$, the energy of a proton machine must be 1800 times larger than an electron machine to reach the same cutoff frequency (not to mention the fact that the magnets have to be 1800 times stronger to achieve the same bend radius $\rho$). At the Fermilab Tevatron, where $c/\rho \sim 3\times10^5$ s$^{-1}$, $\gamma$ must be $> 2000$ to obtain significant light in the visible region; however, the Tevatron operates at $\gamma \sim 1000$.

In the late 1970's, Coïsson calculated that there could be an enhancement in the emission spectrum at higher frequencies for light emitted near the edge of a magnet [10,11]. This implied that proton machines might be able to see visible synchrotron radiation at lower energies than previously thought, and indeed, observations were made using the CERN SPS at proton energies of 300 - 400 GeV [12,13]. Since then, synchrotron radiation has been utilized at the Tevatron [14-16], in the proton beam of HERA at DESY [17], and extensively at the LHC [18] where in addition to a dipole source, there is a dedicated wiggler to enable radiation at lower energies.

Fermilab's effort started in the early 1990's at a time when the only transverse profile measurements were obtained from the flying wire systems. The disadvantage of a flying wire device is that it causes an increase in the emittance every time the wires are flown through the beam. Thus the wires were only flown occasionally and there was no continuous profile measurement. Synchrotron radiation provided the means to make a continuous profile measurement and led to the development of a gated camera based imaging system for the light emitted from the edge of a Tevatron superconducting dipole magnet [19].

As a byproduct of the transverse profile research, it was realized that synchrotron radiation might be useful in solving another monitoring issue [20]: how to measure the relatively low levels of unbunched beam in the abort gaps [21], which was expected to be $\sim 10^4$ times less than the regular beam. Beam present in the abort gaps when the abort kicker magnets fire, is deflected in a wide path and can cause magnet quenches and damage to silicon detectors downstream. The advantage of an optical signal in detecting this small amount of beam is that EMF interference is not amplified in a typical photo-multiplication scheme and as such, gains of $10^5$ or $10^6$ are easily achievable with little noise. A system was developed which ultimately used a Hamamatsu fast gated micro-channel plate (MCP) type photomultiplier tube (PMT) to measure the beam intensity within the abort gaps [22].

## 2. Tevatron environment

The Tevatron collides protons and antiprotons with energies of 980 GeV. The protons originate in a 750 keV Cockroft-Walton source and are accelerated to 400 MeV by a linear accelerator. From there they are accelerated in the Booster synchrotron to 8 GeV and then by the Main Injector synchrotron to 150 GeV before injection into the Tevatron. Antiprotons are produced from a target by 120 GeV protons and are collected into the Debuncher ring at 8 GeV where they are partially cooled and then transferred to the Accumulator ring where they are further stochastically cooled until they are transferred to the Recycler ring where they are cooled by an electron cooling system until needed by the Tevatron. See [23] for a more thorough discussion of the Fermilab accelerator complex.

The Tevatron is a circular accelerator 6.3 km in circumference and normally collides 36 bunches of protons with 36 bunches of antiprotons at an energy of 980 GeV per beam. They circulate in the same beampipe since they have opposite charge, and are kept apart by electrostatic separators which put them on helical paths. The Tevatron has 1113 rf buckets, where the filled buckets are organized into 3 trains of 12 with an abort gap between each train (figure 2). The rf buckets at 980 GeV have a period of 18.831 ns.

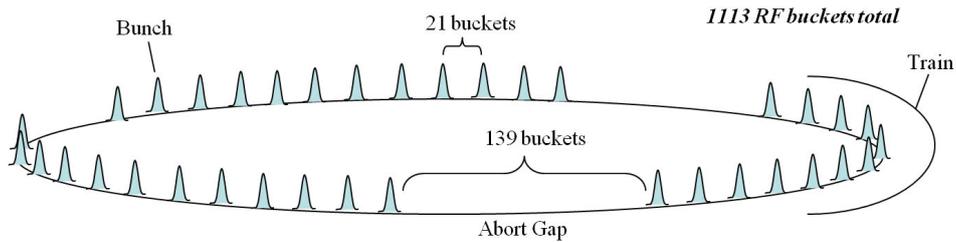

**Figure 2.** Tevatron bunch structure showing the 3 trains of 12 bunches separated by 2.6 µs abort gaps.

## 3. Synchrotron radiation at the Tevatron

The intensity of synchrotron radiation is proportional to $\gamma^4$ and thus increases dramatically with increasing beam energy. For frequencies just above the cutoff, the intensity increases even more dramatically with energy due to the cutoff frequency increasing as $\gamma^3$. At the Tevatron, the cutoff wavelength for radiation from the body of the superconducting dipoles is ~2 µm. Figure 3 shows the strong dependence of synchrotron radiation intensity on beam energy for the Tevatron as calculated by Synchrotron Radiation Workshop (SRW) [24,25].

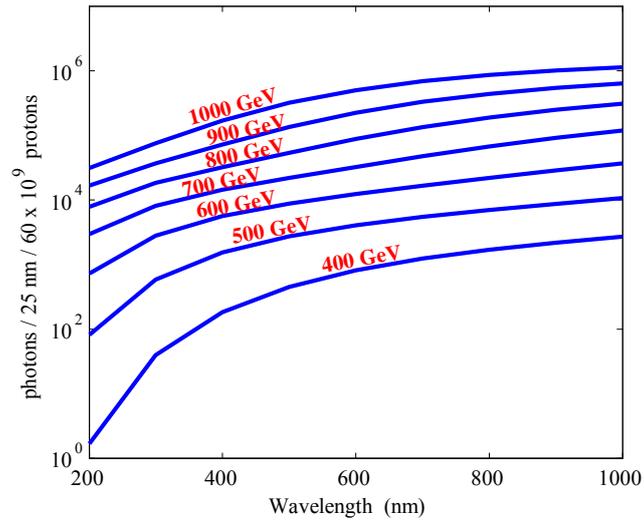

**Figure 3.** Synchrotron radiation spectrum at various proton energies. The Tevatron operates at 980 GeV.

Figure 4 shows the SRW calculated spectrum and transverse distribution of proton synchrotron radiation at the location of the beampipe extraction mirror. The light originates from the far end of a single dipole. One can see how the edge progressively dominates at the shorter wavelengths as evidenced by the single spot.

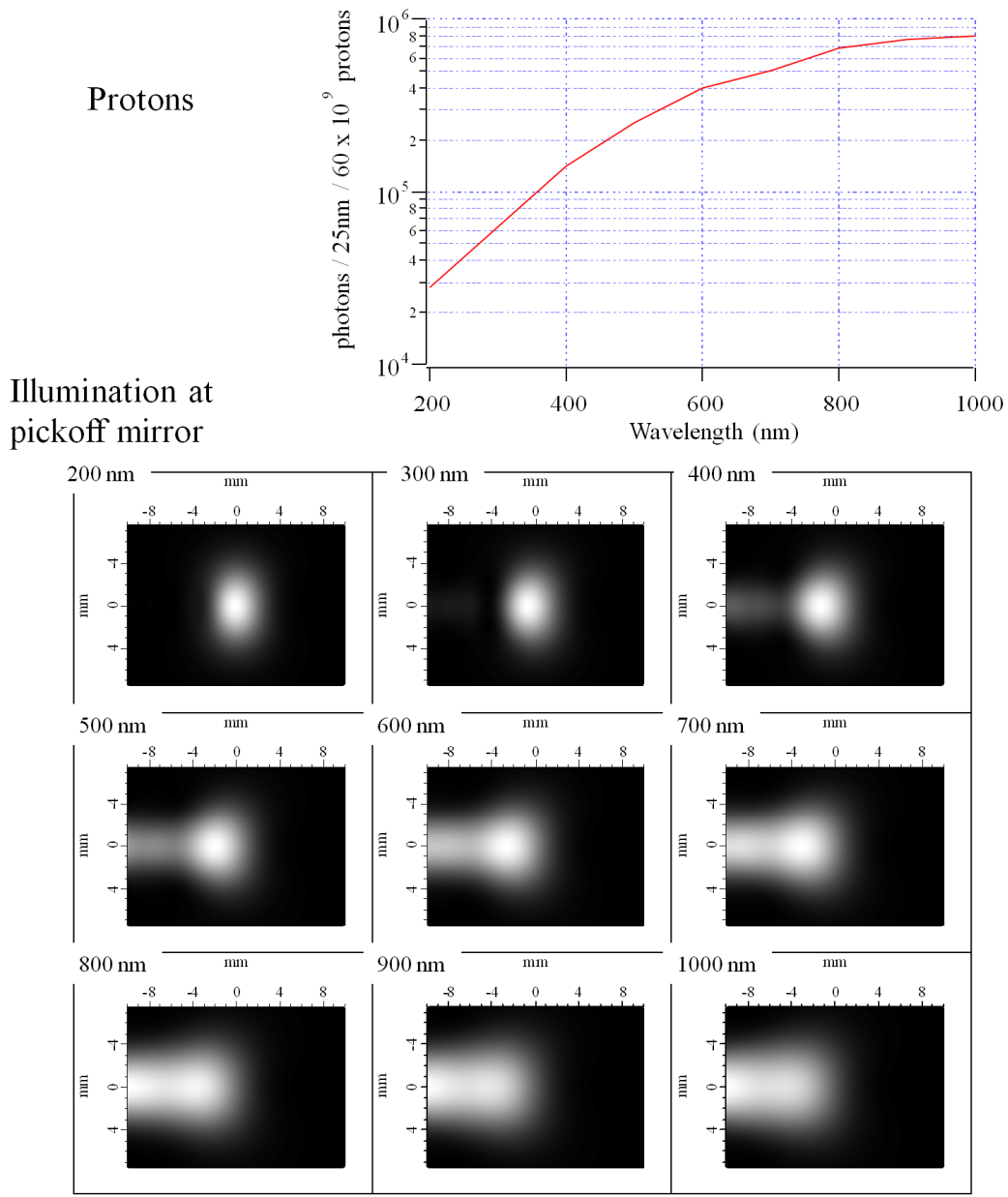

**Figure 4.** Top) Theoretical spectrum of proton synchrotron radiation from the edge of a dipole magnet as a function of wavelength. Bottom) Theoretical spatial distribution of radiation at the point of optical extraction from the beamline for a range of radiation wavelengths. All calculations were performed in SRW [24,25].

Likewise, figure 5 shows the spectrum and transverse distribution of antiproton synchrotron radiation. Here also, the edge dominates at the shorter wavelengths, but the antiproton light originates from the ends of two dipoles separated by ~40 cm, and as such, one can see the interference between the two dipole edges in the 200 nm image (see figure 7 for a beamline schematic showing the two dipole edges).

Antiprotons

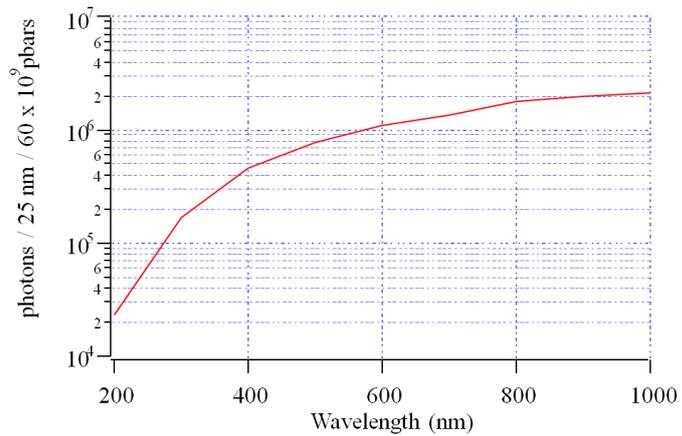

Illumination at
pickoff mirror

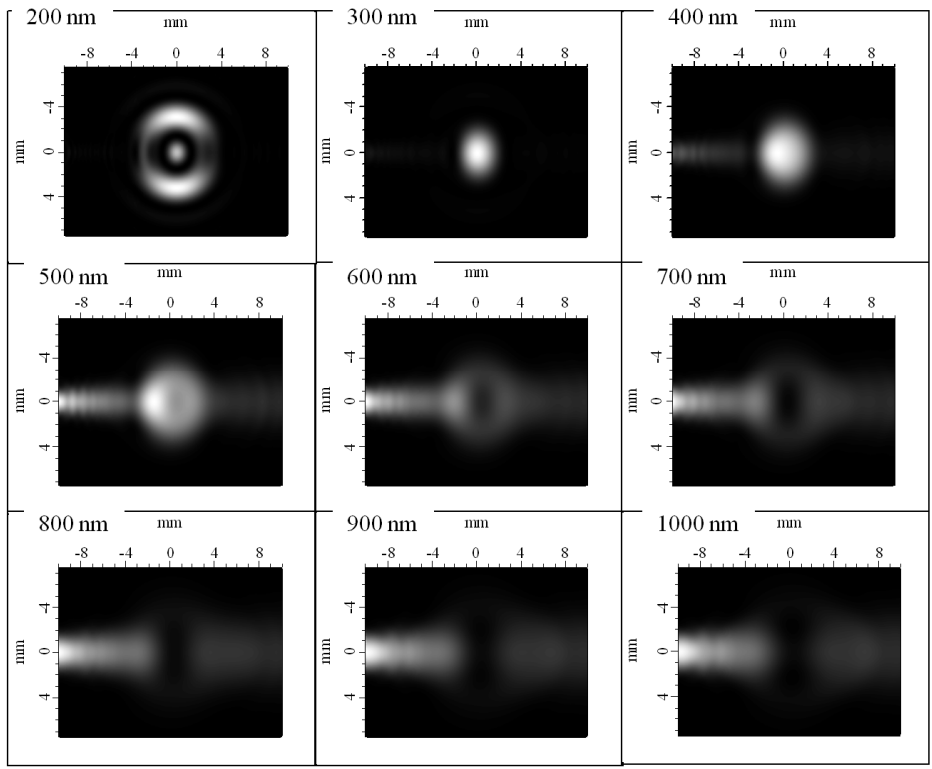

**Figure 5.** Top) Theoretical spectrum of antiproton synchrotron radiation from the adjoining edges of two dipole magnets as a function of wavelength. Bottom) Theoretical spatial distribution of radiation at the point of optical extraction from the beamline for a range of radiation wavelengths. All calculations were performed in SRW [24,25].

Figure 6 shows the horizontal and vertical projections of the radiation indicating the contribution of the light from the body. This body light can smear the image of the beam since it originates from different depths and thus may be out of focus. The body light in the antiproton case is less because the central spot has contributions from two edges which are partially coherent at the wavelength used by the synchrotron light monitor.

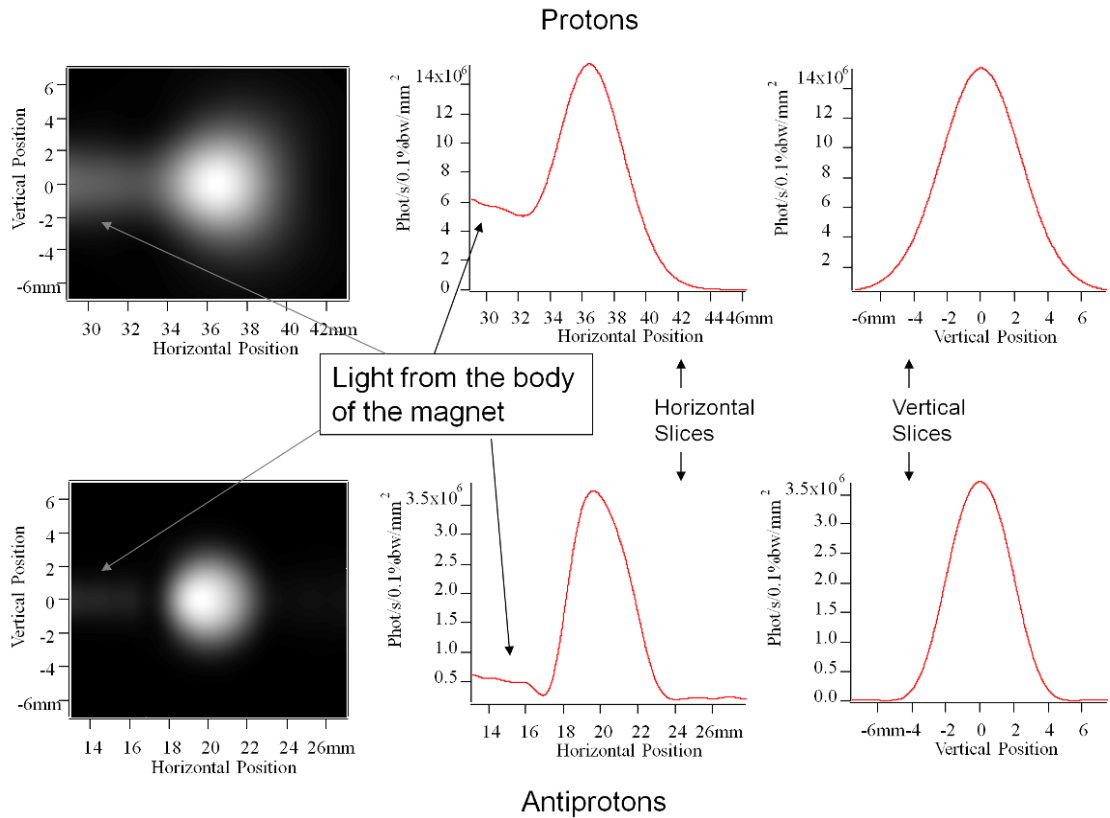

**Figure 6.** Horizontal and vertical projections of the theoretical radiation distribution for the proton and antiproton configuration at the extraction mirror, where the spot size has an rms of ~2 mm.

## 4. Monitoring devices

There are two types of devices that make use of synchrotron radiation in the Tevatron: a transverse profiling device, called Synclite, which images the beam, and an abort gap beam intensity monitor, called AGI, both of which are housed in a single optical system. There are separate optical systems for proton and antiproton beams collocated in a ~3 m warm section between a full-length superconducting dipole magnet and a half-length superconducting dipole magnet, just downstream of the C0 region (figure 7). This is the only place in the Tevatron where a warm section is close enough to both an upstream and downstream dipole to allow extraction of both proton and antiproton light before it encounters the beampipe.

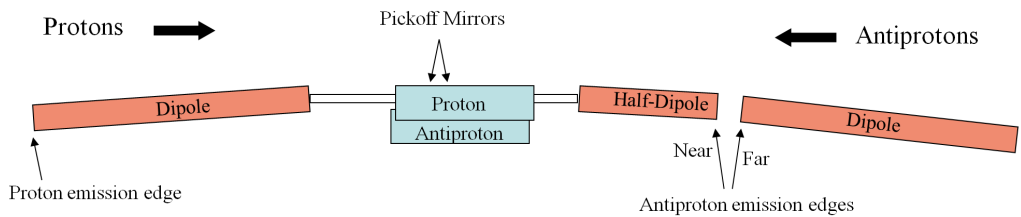

**Figure 7.** Layout of the synchrotron radiation monitor in the Tevatron beamline. Both the half-dipole and the full dipole edges contribute to the radiation. The distance from the pickoff mirror to the proton(antiproton) edge is ~7(5) m. The distance between the two dipole edges in the antiproton case is ~0.4 m.

Each optical setup consists of an aluminized extraction mirror mounted on a motorized insertion stage which directs the light out of the beampipe through a quartz vacuum window to a light tight box (figure 8). The quartz window for the proton(antiproton) setup is an MDC Vacuum Products #450024(#450023). Within this box, the light is first focused by a lens and then redirected by an Al mirror to a Thorlabs #BS013 nonpolarizing beamsplitter. The proton(antiproton) focusing lens is an Oriel 50 mm DIA plano convex lens #40825(#40815) made of BK7 glass with a 1500(750) mm focal length. Just before the splitter is a neutral density filter that can be inserted in the optical path and is used for calibrating the abort gap beam photomultiplier tube (PMT) and occasional studies of the system. The light transmitted through the beamsplitter traverses a blue filter and is received by an intensified camera. In the proton box the blue filter is a Thorlabs FB440-10, with a 440 nm central wavelength and 10 nm bandwidth, and is mounted in a filter wheel with multiple filters for studying diffraction effects. The antiproton box has just a single Melles-Griot 03-FIV-026, with 400 nm central wavelength and 40 nm bandwidth. These filters are necessary to isolate the wavelengths where the edge radiation is most enhanced (see figures 4 and 5). The difference in filter bandwidths between proton and antiproton systems is a byproduct of using a collection of filters for the proton system and not for any particular purpose. The intensified camera is a combination of a Hamamatsu V6887U-02 gated image intensifier with a maximum gain of ~1000 and a minimum gate width of 50 ns, and a Thermo Scientific CID3710DX12, which is a monochrome charge-integrating device (CID) solid state camera with 11.5 μm x 11.5 μm pixels. The two devices are coupled through fiber optic plates. CID cameras are used since they are more radiation tolerant than charge-coupled device (CCD) cameras.

For the abort gap beam intensity measurement, the reflected light from the beamsplitter passes through another neutral density filter, which is also used for calibration purposes, before ending up at the microchannel plate type PMT, a modified Hamamatsu R5916U-50 (built with 3 stages of microchannel plates) with a maximum gain of ~$10^7$ and a minimum gate width of 5 ns.

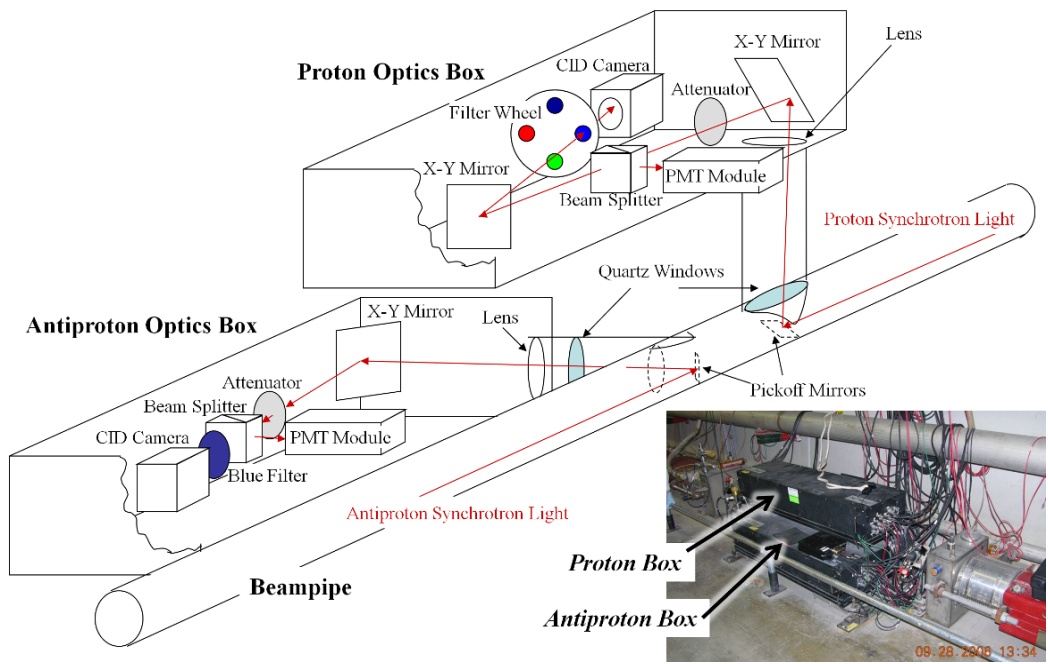

**Figure 8.** Three-dimensional schematic of optical transport lines. The 'PMT Modules' are photomultiplier tubes for the abort gap monitoring devices. The filter wheel in the proton box was installed for diffraction studies but is normally set to a 440 nm center, 10 nm bandwidth filter. The picture is of the two light tight boxes in the accelerator tunnel. The red object on the right side of the picture is the end of the half-dipole magnet.

## 5. Transverse profile monitor

The transverse profile monitor, a.k.a. Synclite, is a single lens imaging system. The proton(antiproton) object and image distances are 769(503) cm and 187(85) cm giving an optical magnification of 0.24(0.17). The distances for the antiprotons are measured to the near source (see figure 7). Table 1 lists the number of photons expected per bunch after each optical element assuming a proton(antiproton) bunch intensity of 250(100) x $10^9$.

**Table 1.** Number of expected photons after the elements in the optical path starting with the generated number at the magnet edge. The generated photon number is per bunch which is ~250 x $10^9$ for protons and ~100 x $10^9$ for antiprotons. The last row indicates the number of photoelectrons. The fact that the number of starting photons is the same is somewhat coincidental. From figures 4 and 5 one can see that the radiation intensity for the antiprotons (at 400 nm) is ~2.5 times that of the protons (at 440 nm), whereas the bunch intensity for protons is ~2.5 times that of antiprotons, thus cancelling.

| After object… | Object Efficiency | | # of photons / bunch / 25 nm | |
|---|---|---|---|---|
| | | | Protons | Antiprotons |
| Magnet Edge | — | | 750 000 | 750 000 |
| Pickoff Mirror | 90% | | 675 000 | 675 000 |
| Vacuum Window | 90% | | 608 000 | 608 000 |
| Lens | 93% | | 565 000 | 565 000 |
| Al Mirror | 90% | | 509 000 | 509 000 |
| Beam Splitter | 44% | | 224 000 | 224 000 |
| Al Mirror (proton only) | 90% | | 202 000 | — |
| Wavelength Filter | 10 nm 40% | 40 nm 160% | 81 000 | 358 000 |
| Photocathode | 14% | | 11 000 p.e. | 50 000 p.e. |

## 5.1 Data acquisition system

Figure 9 shows a layout of the data acquisition (DAQ) system for Synclite. The output of the CID camera is processed by the controller which sends an analog RS-170 video signal out of the tunnel to a National Instruments PCI-1409 10-bit framegrabber card in a Windows XP PC running LabVIEW. The image intensifier is gated using a 150 V pulse from a custom FET pulse amplifier. The pulsing is controlled by a VME based board (VRFT Timing Card) which is synchronized to the Tevatron rf system. Tevatron state information is obtained from another VME based board (UCD Clock Decoder) which decodes the machine state information and allows data taking to happen only when the beam is at sufficient energy to produce measureable synchrotron radiation.

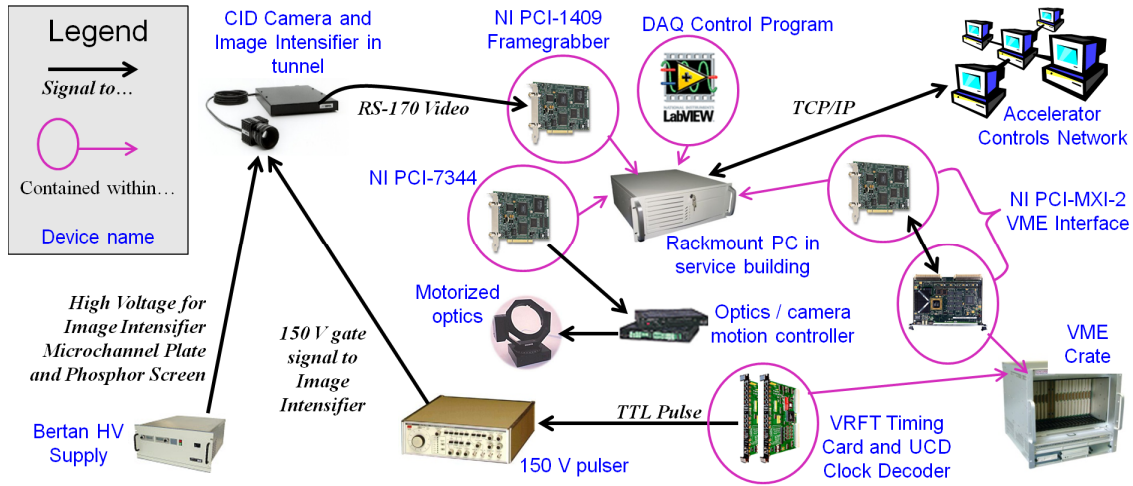

**Figure 9.** Data acquisition layout of Synclite. A LabVIEW program controls the image acquisition and timing of the gates to the image intensifier. It also controls the raising and lowering of the intensifier high voltages. The motorized optics and camera mount in the optics box are controlled separately from the normal data acquisition and are adjusted only occasionally when the beam position moves significantly.

The DAQ is setup to make a measurement of one bunch at a time, with the intensifier being gated on when the desired bunch is present. The number of beam revolutions between successive intensifier gates is controlled by the program to keep the light intensity on the camera within the center region of the dynamic range of the camera. Since a frame is 30 ms and a revolution is 21 µs, there can be a maximum of ~1500 beam 'snapshots' within a video frame. Due to limitations in the gate generator, the minimum number of snapshots in a frame is about 40.

Once 4 video frames have been acquired and averaged, a region of interest is chosen around the peak and both a static background image subtraction and a line by line linear background subtraction are performed on the image. The static background image is acquired at the beginning of a Tevatron store and removes fixed pixel offsets which have an rms of 4-5 camera counts on a baseline of between 60 and 80 counts (the maximum count is 1023). The typical pixel noise after averaging the 4 video frames has an rms of ~14 counts, independent of the intensifier gating duty cycle. The dynamic range of the system is about 1200 (30 for the camera, and 40 from the intensifier gating duty cycle), and the signal to noise ratio at the normal working point is ~500/14.

Using the averaged, background-subtracted images (figure 10), horizontal and vertical projections are made and fitted with a Gaussian plus linear background (figure 11). The fit parameters have various calibrations applied (see below) before they are made available to the accelerator controls network (ACNET). To make a measurement of all 36 proton and 36 antiproton bunches takes approximately 60 seconds.

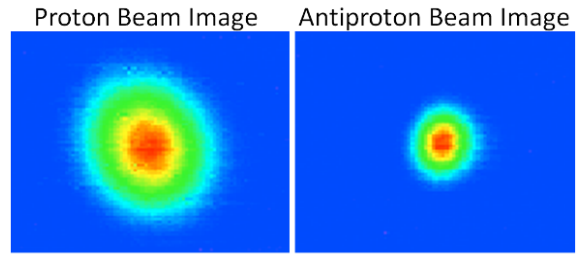

**Figure 10.** Proton and antiproton beam images. The scales are not exactly the same, but in general the antiproton beam is smaller than the proton beam at this location in the accelerator.

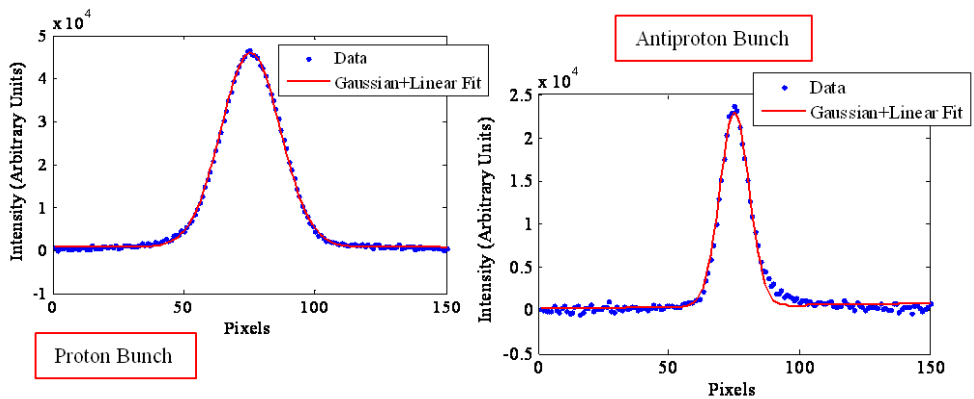

**Figure 11.** Proton and antiproton beam image profiles in the direction of the dipole bend. The light originating from the magnet body is clearly visible as a tail in the antiproton profile. This tail alters the Gaussian fit resulting in a correction that has to be applied to the fitted width.

## 5.2 Transverse profile calibrations and studies

### 5.2.1 Image intensifier and CID camera

Since the goal of this monitor is to measure the beam profile, the variation of the measured sigma was checked with varying levels of light at different locations in the intensified camera chain. Figure 12 shows the results of changing the microchannel plate voltage and the number of beam snapshots per image, and inserting an optical attenuator. The plot demonstrates that the variation in sigma is not caused by any one of these changes. In fact it is most strongly correlated with the light intensity present at the CID camera. This would seem to rule out a non-linearity in the photocathode or the microchannel plate. Possible sources of this include a non-linearity in the phosphor screen of the intensifier or pixel bleeding in the CID camera sensor. Corrections to the measured beam sigma are parameterized as a function of camera intensity and measured sigma (figure 13).

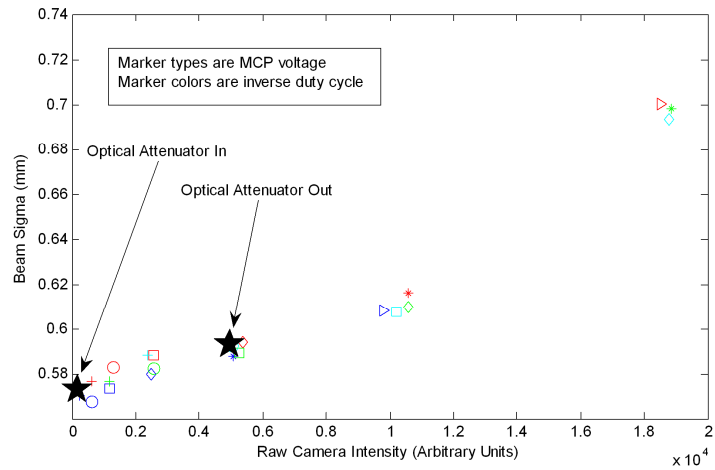

**Figure 12.** Nonlinearity of measured sigma plotted as a function of light intensity at camera. Different markers represent distinct microchannel plate voltages, i.e. intensifier gain, and different colors represent different number of beam pulses per image, i.e. how many times the intensifier was gated on during the camera image.

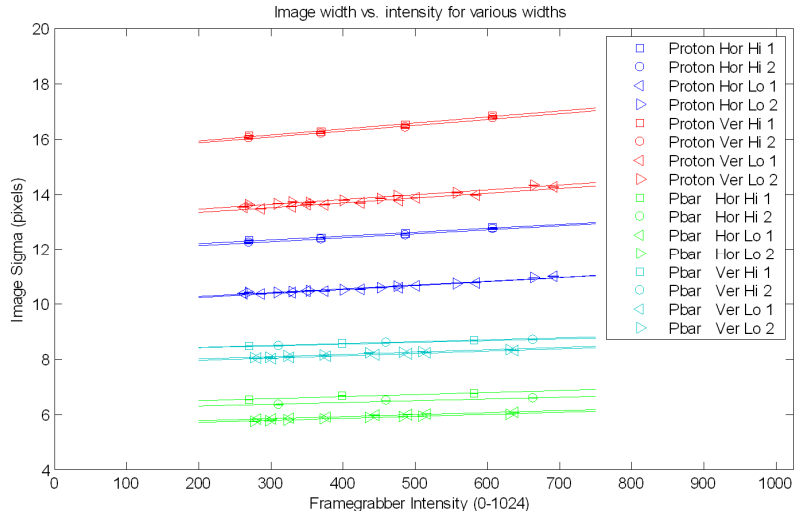

**Figure 13.** Nonlinearity of measured beam sigmas (horizontal and vertical) as a function of camera intensity for both protons and antiprotons of varying bunch intensities.

The distance from a camera to a lens is normally set by minimizing the focused spot size. In this case however, the simulation and data were compared for a range of camera distances (see figure 14). The theoretical distance for the focal length is not where the minimum spot size occurs, presumably because the source is not a zero depth source. For Synclite, the camera position was set to the theoretical focal point as determined from the simulation using this scan. The simulation was then used for corrections due to body light and diffraction (see below).

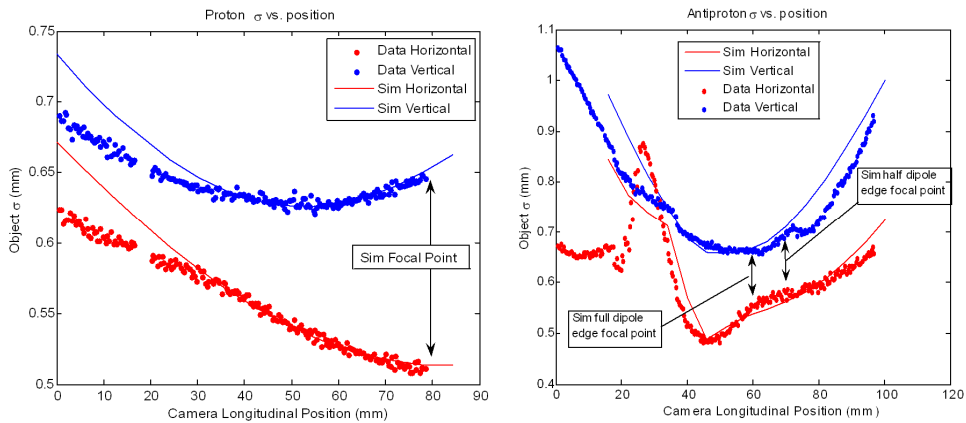

**Figure 14.** Comparison of measured beam size to simulated beam size as a function of the distance to the imaging plane. The boxes indicate the single lens ideal image distance(s) which for the antiprotons are twofold due to the two magnet edges. Image corrections are determined from the simulation based on the location of the camera obtained from this plot. The agreement between simulation and data is fairly good except for regions far from the focal points.

### 5.2.2 Synchrotron radiation generation and optical transport

The beam image can include light originating from the body, and as seen in figure 11, that light can distort the image. To study the generation and transport of the radiation, measurements were performed comparing the simulated beam image to the measured beam image with the extraction mirror located at different positions. With the mirror at different positions in the bend plane, the beam images contain differing amounts of body light, allowing confirmation of the behavior of the optical system as well as providing an absolute measurement of the position of the mirror relative to the beam. Figures 15 and 16 show the results of the proton and antiproton study respectively.

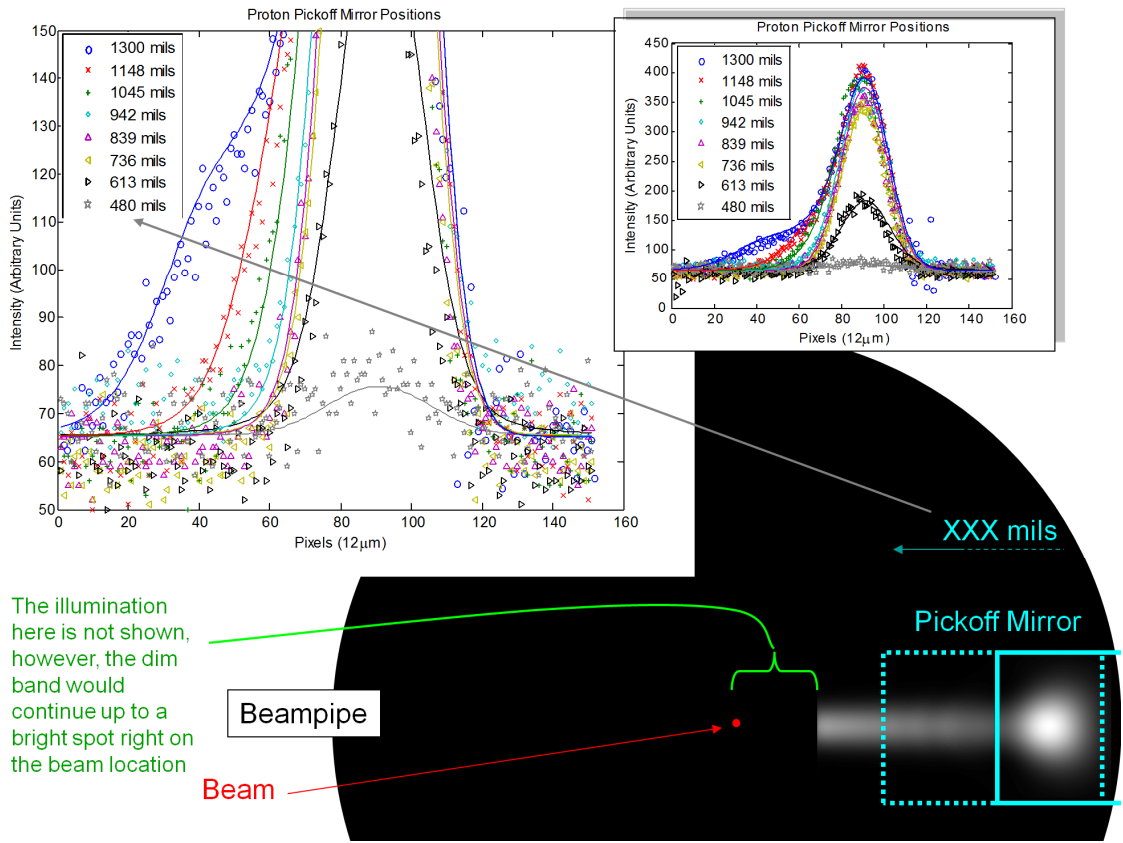

**Figure 15.** Proton study moving the extraction mirror in the bend plane. The diagram shows the relative positions of the proton extraction mirror (solid rectangle is the nominal position) and synchrotron radiation in the beampipe. The plots show the calculated (solid) and measured (points) beam image profiles in the bend plane. The number of mils (1mil = 25.4 µm) represents only a relative position of the pickoff mirror for each step. For reference, the 480 mils position corresponds to the left edge of the mirror being positioned just to the right of the bright spot, and the 1300 mils position corresponds to the dotted rectangle. The inset plot is the same plot as the large one, but shows the full height of the distributions.

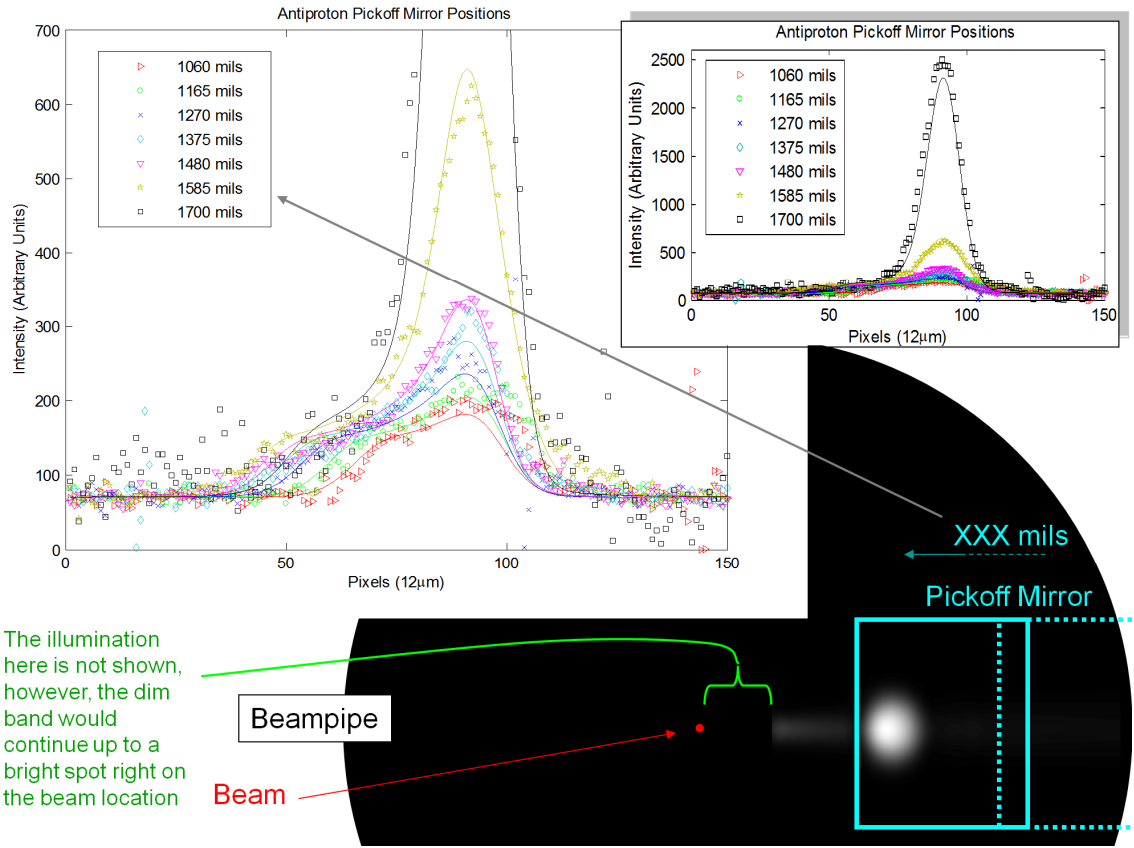

**Figure 16.** Antiproton study moving the extraction mirror in the bend plane. The diagram shows the relative positions of the antiproton extraction mirror (solid rectangle is the nominal position) and synchrotron radiation in the beampipe. The plots show the calculated (solid) and measured (points) beam image profiles in the bend plane. The number of mils (1mil = 25.4 μm) represents only a relative position of the pickoff mirror for each step. For reference, the 1700 mils position corresponds to the left edge of the mirror being positioned on the left edge of the light spot, and the 1060 mils position corresponds to the dotted rectangle. The inset plot is the same plot as the large one, but shows the full height of the distributions. Note that despite the fact that it cannot be easily seen in the diagram above, there is also a band of light on the right side of the spot which is seen in the plots.

### 5.2.3 Diffraction

Since the opening angle of the synchrotron radiation is proportional to $1/\gamma$, $\gamma$ being the Lorentz factor (1044 at the Tevatron), diffraction can play a significant role in the beam image. The well known microscope diffraction formula, i.e. the radius of the airy disk, is

$$r_{airy} = 0.61 \frac{\lambda}{NA} \qquad (5.1)$$

where $NA$ is the numerical aperture of the objective lens ($NA = n\, sin\theta$; where $n$ is the refractive index of the medium before the lens), and $\lambda$ is wavelength. For Synclite, $NA$ is roughly the half angle of the light cone (since the optical elements are all much bigger than the light cone) which is 0.001 (=$1/\gamma$) for the vertical plane and 0.0005 (=$1/(2\gamma)$) for the horizontal bend plane If one takes a more Gaussian approach, and says that the diffraction sigma is 1/2 the airy radius, then at 400 nm, $\sigma_{vertical}$ = 120 μm and $\sigma_{horizontal}$ = 240 μm.

The filter wheel in the Synclite proton box has 4 wavelength filters (10 nm bandwidth): 360 nm, 440 nm, 530 nm, and 620 nm. Measurements of the vertical beam sigma were taken with each of the filters at the nominal focal point for that wavelength. Figure 17 shows the variation of beam sigma with wavelength for both data and simulation and also shows the functional behavior of the microscope formula.

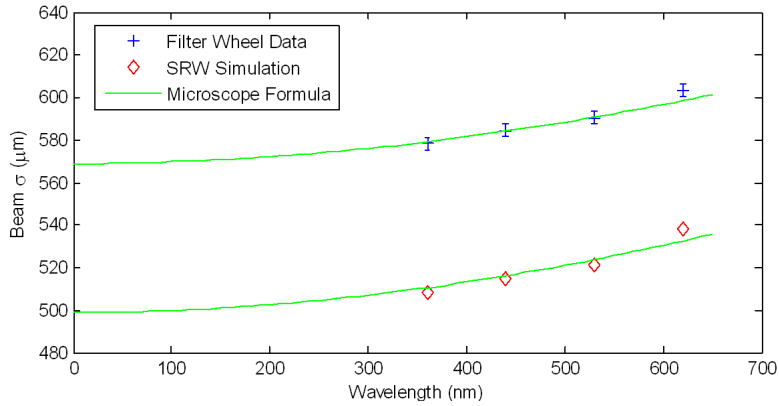

**Figure 17.** Comparison of the measured vertical beam sigmas from data with simulation and with the microscope formula. The shape of the data agree with the simulation and both agree with the microscope formula. The last point at 620 nm is high due to the fact that the camera could not be moved far enough to reach the minimum.

After correcting for the above nonlinearities, a global resolution term of 2.0-2.5 pixels was estimated by comparing the synchrotron light profiles with profiles from the Tevatron Flying Wire systems. Over the course of a single store, the ratio of the two profiles for a single bunch should remain constant. The resolution term was varied until the ratio was constant. This resolution was checked at other times and was found to be unchanged.

### 5.3 Performance

With all the calibrations in place, the linearity of the system was checked using proton bunches of varying intensity. These measurements were taken during a store when there were no antiprotons and only 18 proton bunches with large variation in bunch intensity. The Synclite measured intensity was compared to a measurement by a wall current monitor (figure 18). The agreement is good to within 5%.

Figure 19 shows a comparison of the measured horizontal and vertical emittances of Synclite and the flying wires for both protons and antiprotons. Since the Synclite and flying wire systems are not physically at the same location in the ring, a comparison of profile widths is not meaningful. Instead the emittance, $\varepsilon$, is calculated from the profile widths, $\sigma$, the Tevatron beta functions, $\beta$, the dispersions at the two locations, $D$, the relativistic Lorentz factor, $\gamma$, and the measured momentum spread, $dp/p$.

$$\varepsilon = \frac{6\gamma}{\beta}\left(\sigma^2 - D^2\left(\frac{dp}{p}\right)^2\right) \quad \pi \text{ mm mr} \quad (5.2)$$

The ratio of the two monitors is plotted over the course of about a dozen stores (almost three weeks time) and its deviation from unity is less than 10-15%. This is consistent with expected uncertainties in the Tevatron lattice functions.

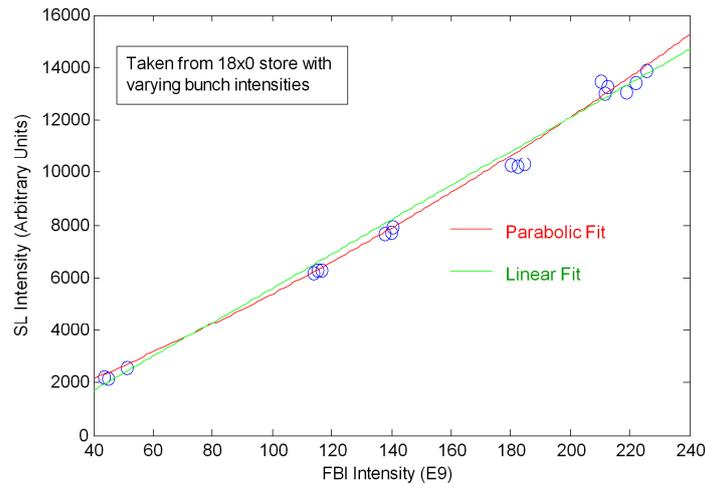

**Figure 18.** Plot of synchrotron radiation intensity (vertical axis) versus wall current measured intensity (horizontal axis) for a collection of proton bunches with varying intensities. The variation from linear is < 5%.

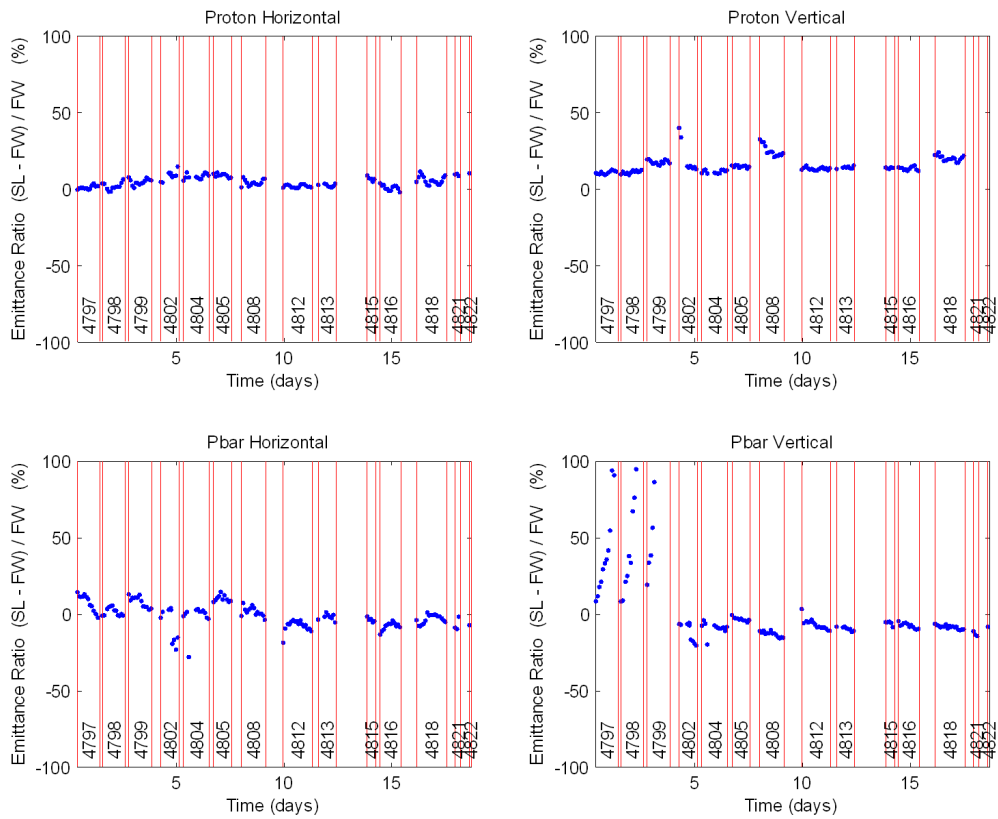

**Figure 19.** Comparison of emittances calculated from Synclite and flying wire data over the course of almost 3 weeks (the number labels in the plot refer to Tevatron store numbers). The agreement is within 10-15% which is within the uncertainties in the Tevatron lattice measurements. The first few sections in the antiproton vertical plot (labeled Pbar Vertical) contained fitting failures due to the image position on the camera.

Figure 20 shows the emittance growth over the course of two stores from 2008 and 2011. The agreement between the Synclite system and the flying wires has become slightly worse over time in part due to degradation of the image intensifier.

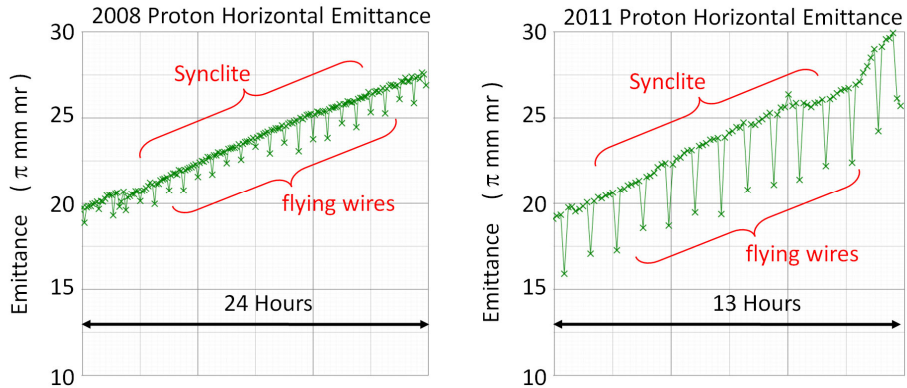

**Figure 20.** Proton horizontal emittances from Synclite (higher points) and the flying wires (lower points) from 2008 and 2011 showing the characteristic increase in the emittance over the course of the store. The discrepancy between the two systems has increased over the last several years in part due to the degradation of the image intensifier. There is a decrease in the emittance near the beginning of the 2008 data due to a change in the betatron tune which scraped part of the beam away. At the end of the 2011 plot, there was a problem with the the Tevatron electron lens which caused the observed increase in emittance growth.

In 2007, the brightness of the antiprotons increased with improvements to the electron cooling in the Recycler, causing proton beam losses and decreased beam lifetime. The solution was to slightly increase the emittance of the antiprotons using a noise-based rf kicker [26]. Figure 21 shows the increase in emittance as measured by Synclite during the time when the kicker is on.

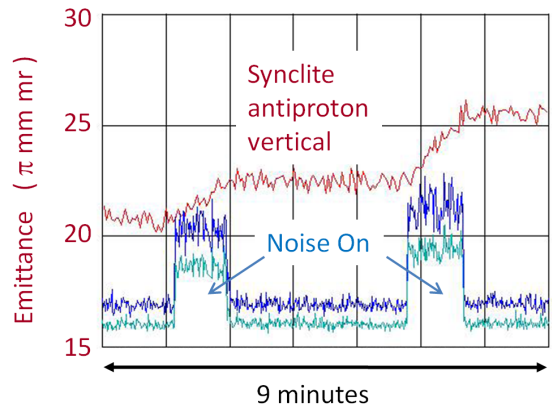

**Figure 21.** Plot showing the increase in antiproton emittance when the noise kicker is on. Synclite was used to determine the emittance growth rate for the kicker. This figure modified from figure 4 in [26].

### 5.4 Operational use

Synclite became an essential tool at the very early stages of commissioning of the Tevatron Collider Run II, and contributed significantly to the understanding and modeling of beam dynamics effects. It was particularly useful in understanding the bunch-to-bunch differences caused by long-range beam-beam interactions [27]. Synclite was one of the primary instruments providing data for the development of luminosity evolution modeling and optimization.

Since the very beginning of Run II, Tevatron performance was significantly limited by the effects of beam-beam interactions. The effects included increased levels of particle losses and rapid emittance growth when the beams were brought into collision. The latter effect was predominantly observed in the antiproton beam and at times resulted in as much as 20% reduction of the collider luminosity. With the use of Synclite, it was demonstrated that the emittance blow-up happens on the time scale of one minute and depends strongly on the closeness of the betatron tune working point to the fifth-integer resonance ($Q = 0.6$). Theoretical studies suggested that this phenomenon must be caused by the combination of long-range and head-on beam-beam interactions, and should result in a distinct bunch-to-bunch difference in the emittance growth. Again, Synclite was instrumental in demonstrating that this was indeed the case (figure 23 in [27]).

Once the origin of the antiproton emittance blow-up was realized, Synclite became the tool for the routine monitoring of the "scallop" effect and directed the tune working point adjustments. Ultimately, an improved beam separation scheme was implemented in 2006-2007 [28], which allowed an increase in the beam brightness and consequently an increase in the collider luminosity.

The second important application of the instrument was for the development of a model of the collider luminosity evolution. For the optimization of the machine luminosity performance, it was essential to accurately model the evolution of beam parameters including intensity, luminosity, and transverse and longitudinal beam sizes during a collider store. Many beam dynamics processes govern the behavior of these quantities, and building a model describing all effects was a nontrivial physics task. Such a model was built [29] and benchmarked against the observations, of which Synclite was an essential component. Ultimately, a computer application program was put online, analyzing the machine performance in every high energy physics store [30].

During the last years of the collider run, Synclite operated in two modes during a collider store. First, only two bunches from each beam were monitored, allowing for fast measurements during the critical stages of the collider store – the so-called squeeze and initiate collisions phases, when the transverse beam dimensions and positions are changing rapidly. Then, the device was put in the mode monitoring all 36 bunches in both beams for the remainder of the store. Transverse beam size measurement played an important role in a number of beam physics experiments, including but not limited to the studies of beam-beam effects and their compensation with electron lenses [31] and hollow electron beam collimation [32].

## 6. Abort gap beam intensity monitor

The Abort Gap Beam Intensity Monitor, a.k.a. AGI, is designed to monitor the intensity of the beam present in the Tevatron abort gaps. In particular, the second abort gap which is where the abort kicker magnet fires. It consists of a photomultiplier tube (PMT) which measures the intensity of the synchrotron radiation. Table 2 lists the amount of light present at various points in the system and at various wavelengths.

**Table 2.** Numbers of photons after elements in the optical detection path for various wavelengths. The numbers are per 100 nm bandwidth per $10^9$ particles evenly distributed around the ring. The efficiencies for optical elements before the beamsplitter can be found in Table 1. The photon numbers after the quantum efficiency of the photocathode are of course photoelectrons. The duty cycle represents the length of the abort gap relative to the total ring.

| After Object… | # of photons / 100 nm / $10^9$ particles | | | | | | | |
|---|---|---|---|---|---|---|---|---|
| | Protons | | | | Antiprotons | | | |
| **Wavelength (nm)** | **450** | **550** | **650** | **750** | **450** | **550** | **650** | **750** |
| Magnet Edge | 15 000 | 24 000 | 30 000 | 37 000 | 39 000 | 58 000 | 80 000 | 106 000 |
| Optics efficiency through beamsplitter (30%) | 4 500 | 7 200 | 9 000 | 11 100 | 11 700 | 17 400 | 24 000 | 31 800 |
| Quantum Efficiency (13%,8.5%,5.5%,2.0%) | 590 | 610 | 490 | 220 | 1 520 | 1 480 | 1 320 | 640 |
| Duty Cycle (10%) | 191 | | | | 496 | | | |

### 6.1 Data acquisition system

The DAQ system (figure 22) consists of a 5-slot VME crate with an MVME 2434 crate processor running the VxWorks operating system from WindRiver, a COMET 12-bit digitizer board, a VME-based timing board (VRFT Timing Card) for beam timing, and a fast gated charge integrator with a 250 Ω input impedance and an output calibration of ~12 V/nC. The timing gates for the PMT, integrator, and digitizer, are generated by the VRFT board from beam synchronization events and delayed appropriately via a NIM delay module. The PMT requires its gates to be greater than 10 Volts, so the timing signals generated by the VRFT are passed through a TTL to 15 V amplifier. This amplifier, along with a high voltage power supply and the delay module, reside in a NIM crate. The PMT anode signal is brought out of the tunnel to the integrator which feeds the digitizer board. The digitizer board is read out by the DAQ program running on the crate processor. Each measurement consists of an average of 1000 readings over several milliseconds and includes pedestal measurements interspersed with the signals.

There are three separate measurements, one in each of the three abort gaps. The measurements do not span the entire abort gap however. They cover about 20% of each gap and are located at different positions in each gap. The most important one of these is the measurement in abort gap two where the abort kicker is fired. This measurement occurs early in the abort gap during the time that the abort kicker field would still be rising to its flattop value. It is during this time that any beam present would be kicked into the CDF silicon detector causing damage to the detector. The measurements in the other abort gaps are placed in the middle of the gap and at the end of the gap. This spacing provides a crude sampling of the abort gap beam distribution assuming that all three gaps have roughly the same distribution.

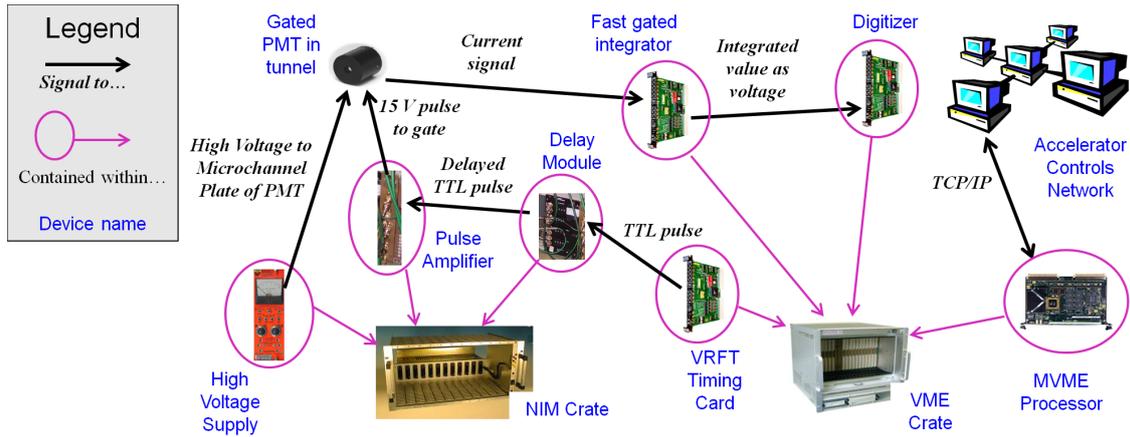

**Figure 22.** Schematic of data acquisition system for the abort gap beam intensity monitor. The PMT requires high voltage for the microchannel plate, and a low voltage (15 V) gate.

### 6.2 Calibrations

#### 6.2.1 Pedestal measurement

There are two pedestal subtractions performed on the PMT data. The first is a static pedestal subtraction. The static pedestal determination is done manually by inserting the neutral density filters and taking a measurement in the abort gap. It is preferable to do this while beam is in the Tevatron since EMF noise conditions change depending on the state of the accelerator. The second pedestal is a dynamic one measured at the same time during the beam revolution except 3 turns later. This measurement is accomplished by not gating the PMT and as such contains microchannel plate dark currents, EMF noise, and temperature variations in the integrator. The dynamic pedestal adjusts the static pedestal for drifts over time. Figure 23 shows the drift in the baseline with both previous lower-gain PMT and pedestal subtraction algorithm and the new higher-gain PMT and improved pedestal subtraction algorithm. The latest measurements indicate a stability of the baseline of $\sim 10^7$ over a period of 5 months. From this plot the sensitivity of the proton abort gap monitor is estimated to be $\sim 10^7$ particles evenly distributed over the ring. The typical beam intensity in the abort gap is $\sim 10^9$ particles so the signal to noise ratio is about 200. The dynamic range of the system depends on the PMT gain. With the present settings, the maximum signal is $80 \times 10^9$ particles giving a dynamic range of $1.6 \times 10^4$. This is greater than the 10-bit digitizer range due to the averaging of 1000 readings.

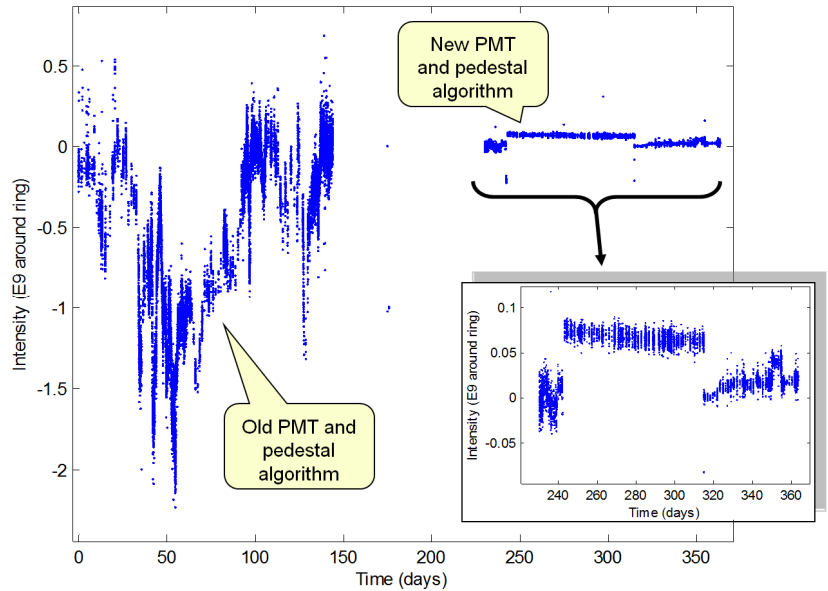

**Figure 23.** Pedestal drift over the course of a year including both the old lower-gain PMT and old pedestal algorithm, and the new higher-gain PMT and new pedestal algorithm. The stability after the improvements is better than $10^8$ unbunched beam particles compared to $>10^9$ before.

### 6.2.2 Gain calibration

The gain of the system can be calibrated from a proton/antiproton bunch by inserting the neutral density filters and moving the gate such that it coincides with a bunch. The neutral density filters had to be calibrated since the spectral response of the PMT extends to the near infrared where the nominal filter value is no longer correct.

An additional calibration study has been done by turning off the Tevatron Electron Lens (TEL) and observing the increase in the abort gap beam (figure 24). The increase measured by the AGI system is compared with the baseline subtracted beam intensity as measured by a DC current transformer (DCCT) which provides a high resolution measurement of the total beam intensity. The quadratic baseline is determined from a period of ~180 minutes before the TEL is turned off. The agreement of the DCCT measurement with the AGI measurement is quite good indicating that the bunch based calibration is accurate. Reference [21] contains detailed information about the generation and monitoring of uncaptured beam in the Tevatron.

One can also look at the abort gap intensity of antiprotons during the TEL study (figure 25). The increase in uncaptured antiprotons is observable, but at a level of ~$10^3$ less than the proton intensity.

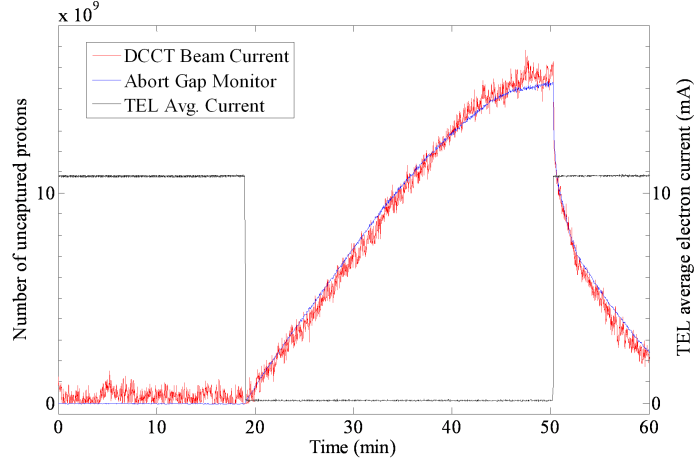

**Figure 24.** Comparison of DCCT measurement of unbunched protons to the AGI measurement. When the TEL current is zero, unbunched beam accumulates in the abort gaps. When the TEL turns back on, the beam is cleared out. This figure is reproduced from figure 15 in [21].

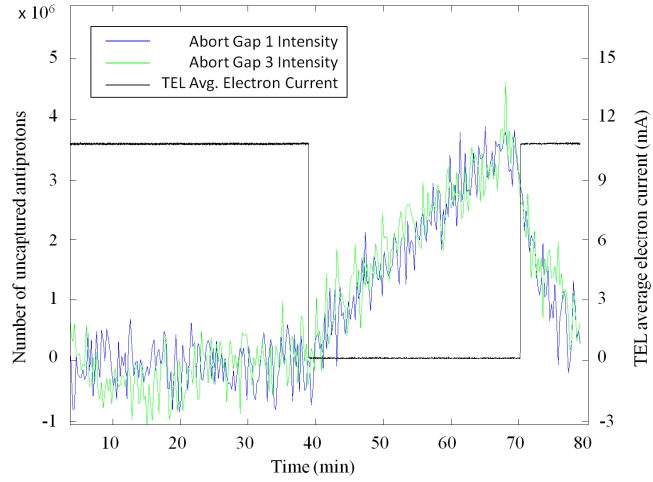

**Figure 25.** Antiproton abort gap levels in gaps 1 and 3 during the TEL study. The levels are $10^3$ less than the proton levels, but they are still visible. From this plot, one can deduce the sensitivity fof the antiproton monitor to be $\sim 10^6$ particles evenly distributed over the ring.

As an additional check of the calibrations, one can compare the proton and antiproton systems given the operating voltages, operating scale factors after calibration, and expected optical luminosities as calculated above. If the calibration is correct, the following should be true.

$$\frac{I_P}{I_A} \times \frac{G_P}{G_A} \times \frac{SF_P}{SF_A} = 1 \qquad (6.1)$$

Here $I$ is the expected synchrotron radiation intensity per particle, $G$ is the gain of the PMT, and $SF$ is the scale factor determined by the calibration ($P$ and $A$ indicate proton and antiproton). With the gains obtained from the gain vs. voltage calibration curve of the PMTs, and the expected intensity ratio of 1 / 2.7 obtained from SRW calculations, the equation works out to a value of 0.92 which is in pretty good agreement.

## 6.3 Performance and operational use

Figure 26 shows the typical abort gap beam intensity just after acceleration. There is a combination of captured and uncaptured beam in the abort gap which is seen by the AGI. Shortly after the beam reaches maximum energy, the TEL is turned on and one can see the measurement that overlaps with the TEL pulse drops rapidly to zero.

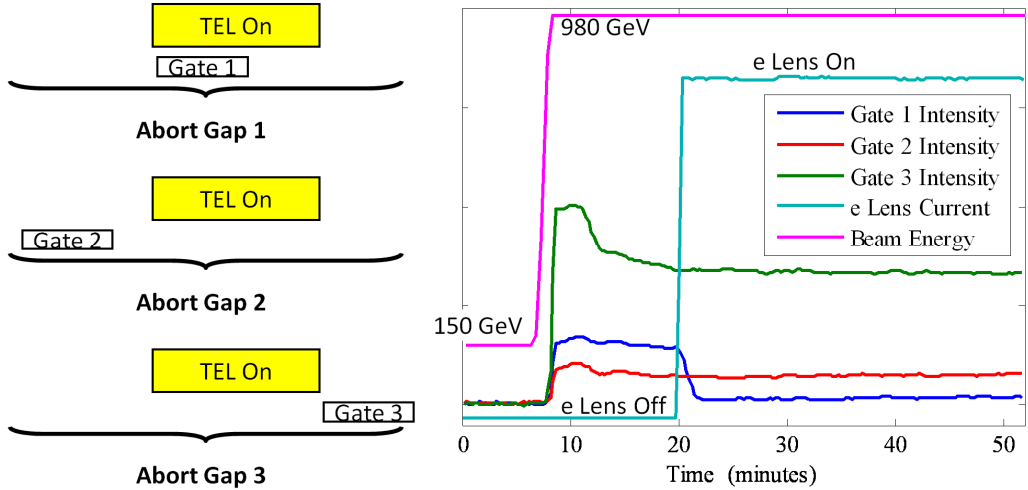

**Figure 26.** Abort gap beam intensity as measured by the AGI at three different times. The left plot is a diagram of the relative timing of various AGI measurements with the TEL. Gate 1 is in the middle of the abort gap and coincides with the TEL. Thus any dc beam clearing action by the TEL is first noticed in Gate 1 as can be seen in the plot shortly after ramping to 980 GeV when the TEL (e Lens) is first turned on.

The AGI was designed to monitor the abort gap beam intensity to prevent magnet quenches and damage to the silicon detector of the Collider Detector at Fermilab (CDF). CDF is located in a more susceptible location than the other Tevatron experiment, DØ. To counter this threat, the monitor is used in several ways. On the accelerator side, the monitor generates a software alarm that is seen by operators who can then respond by manipulating the TEL (see [21] for details about uncaptured beam control). From CDF's perspective, the damage to the silicon is much greater if the detector is on and operating when hit by the stray proton beam so the AGI reading is one of the inputs to a trip system which turns off the detector if the abort gap beam intensity gets too large. Figure 27 is a plot of the abort gap beam intensity following a trip of an rf accelerating station in the Tevatron. Beam is longitudinally disrupted due to the trip and exits the rf bucket where it migrates to the abort gap and is eventually cleared by the TEL.

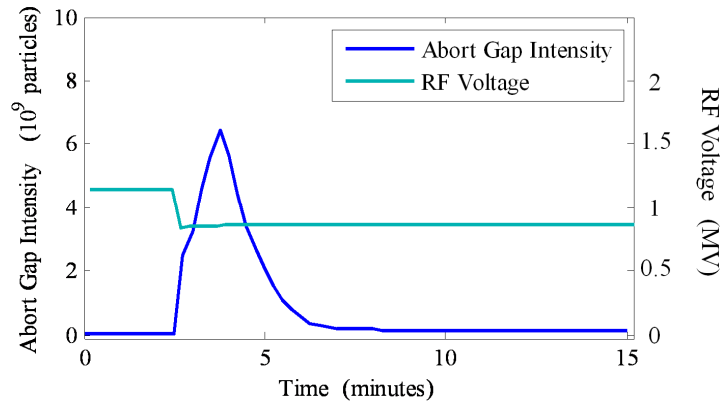

**Figure 27.** AGI measurement following the trip of a Tevatron rf accelerating station. This kind of event would generate an alarm seen by the operators who would then decide what if anything to do. In this case, the normal action of the TEL cleared the dc beam from the abort gap without any further intervention.

## 7. Conclusions

The Synclite and AGI are non-intercepting instruments based on synchrotron radiation from the edge of a Tevatron superconducting dipole magnet. Synclite provides both proton and antiproton bunch-by-bunch beam profiles at a rate of between once every second and once every minute (monitoring fewer bunches corresponds to faster rates). The AGI monitors the beam intensity in the abort gaps once every fifteen seconds to prevent quenches and damage to the experiments from losses when the beam aborts. They have both been extremely reliable and proven to be invaluable tools for optimizing the performance of the Tevatron and protecting it from damage.

## Acknowledgments

The authors would like to thank their colleagues in the accelerator division for their assistance and support in making these devices operational and robust. We would also like to thank Jim Fast (now at Pacific Northwest National Lab) and Ken Schultz for constructing the beamsplitter setup, Heide Schneider and Sten Hansen for the PMT electronics of the first AGI system, John Byrd and Stefano DeSantis for allowing us to use their gated PMT for an extended period of time, and Vladimir Shiltsev for pushing to get these systems operational.

## References

[1] A. Liénard, *Champ Électrique et Magnétique, L'Éclairage Électrique,* **16** (1898) 5.

[2] E. Wiechert, *Elektrodynamische Elementargesetze, Archives Néland* (2) **5** (1900) 549 and reprinted in *Ann. Phys.* **309** (1901) 667.

[3] G.A. Schott, *Electromagnetic Radiation*, Cambridge University Press, Cambridge, U.K. 1912.

[4] I. Pomeranchuk, *On the maximum energy which the primary electrons of cosmic rays can have on the earth's surface due to radiation in the earth's magnetic field, J. Phys. (USSR)* **2** (1940) 65.

[5] D. Iwanenko and I. Pomeranchuk, *On the maximal energy attainable in a betatron*, *Phys. Rev.* **65** (1944) 343.

[6] J. Schwinger, *Electron radiation in high energy accelerators*, *Phys. Rev.* **70** (1946) 798.

[7] J. Schwinger, *On the classical radiation of accelerated electrons*, *Phys. Rev.* **75** (1949) 1912.

[8] F. R. Elder et al., *Radiation from electrons in a synchrotron*, *Phys. Rev.* **71** (1947) 829.

[9] J. D. Jackson, *Classical Electrodynamics, 2$^{nd}$ Ed.*, Equation 14.95, John Wiley and Sons, Inc., New York 1975.

[10] R. Coïsson, *On synchrotron radiation in non-uniform magnetic fields*, *Opt. Commun.* **22** (1977) 135.

[11] R. Coïsson, *Angular-spectral distribution and polarization of synchrotron radiation from a 'short' magnet*, *Phys. Rev.* **A20** (1979) 524.

[12] R. Bossart et al., *Observation of visible synchrotron radiation emitted by a high-energy proton beam at the edge of a magnetic field*, *Nucl. Inst. and Meth.* **164** (1979) 375.

[13] R. Bossart, J. Bosser, L. Burnod, E. D'Amico, G. Ferioli, J. Mann, and F. Meot, *Proton beam profile measurements with synchrotron light*, *Nucl. Inst. and Meth.* **184** (1981) 349.

[14] A. A. Hahn and P. Hurh, *Results from a prototype beam monitor in the tevatron using synchrotron light,* in proceedings of *1991 Particle Accelerator Conference,* p. 1177, 6 – 9 May 1991, San Francisco, California, USA.

[15] A. A. Hahn and P. Hurh, *Results from an imaging beam monitor in the Tevatron using synchrotron light,* in proceedings of *15$^{th}$ Int. Conf. on High Energy Accelerators,* 20 – 24 July 1992, Hamburg, Germany; available at http://beamdocs.fnal.gov/AD-public/DocDB/ShowDocument?docid=185.

[16] H. W. K. Cheung, A. Hahn, and A. Xiao, *Performance of a beam monitor in the Fermilab Tevatron using synchrotron light,* in proceedings of *2003 Particle Accelerator Conference (PAC03),* p. 2488, 12 — 16 May 2003, Portland, Oregon, USA.

[17] G. Kube et al., *Proton synchrotron radiation diagnostics at Hera,* in proceedings of *Beam Instrumentation Workshop (BIW06),* p. 374, 1 – 4 May 2006, Fermilab, Batavia, Illinois, USA.

[18] T. Lefevre et al., *First beam measurements with the LHC synchrotron light monitors*, in proceedings of *International Particle Accelerator Conference (IPAC 10),* 23 – 28 May 2010, Kyoto, Japan.

[19] R. Thurman-Keup, *Proton synchrotron radiation at Fermilab*, in proceedings of *Beam Instrumentation Workshop (BIW06),* p. 364, 1 – 4 May 2006, Fermilab, Batavia, Illinois, USA.

[20] J.-F. Beche, et al., *Developement of an abort gap monitor for high-energy proton rings,* in proceedings of *Beam Instrumentation Workshop (BIW04),* p. 105, 3 – 6 May 2004, Knoxville, Tennessee, USA.

[21] Xiao-Long Zhang et al., *Generation and diagnostics of uncaptured beam in the Fermilab Tevatron and its control by electron lenses*, *Phys. Rev. ST Accel. Beams* **11** (2008) 051002.

[22] R. Thurman-Keup, et al., *Measurement of the intensity of the beam in the abort gap at the Tevatron utilizing synchrotron light*, in proceedings of *Particle Accelerator Conference (PAC 05),* p. 2440, 16 – 20 May 2005, Knoxville, Tennessee, USA.

[23] S. Holmes, R.S. Moore, V. Shiltsev, *Overview of the Tevatron collider complex: goals, operations and performance*, 2011 *JINST* **6** T08001.

[24] O. Chubar and P. Elleaume, *Accurate and efficient computation of synchrotron radiation in the near field region*, in proceedings of *1998 European Particle Accelerator Conference (EPAC98)*, p. 1177, 22 – 26 June 1998, Stockholm, Sweden.

[25] O. Chubar, *Precise computation of electron-beam radiation in nonuniform magnetic fields as a tool for beam diagnostics, Rev. Sci. Instrum.* **66** (1995) 1872.

[26] C. Y. Tan, J. Steimel, *Controlled emittance blow up in the Tevatron*, in proceedings of *Particle Accelerator Conference (PAC09)*, 4 – 8 May 2009, Vancouver, British Columbia, Canada.

[27] V. Shiltsev, et al, *Beam-beam effects in the Tevatron*, *Phys. Rev. ST Accel. Beams* **8** (2005) 101001.

[28] Y. Alexahin, *Optimization of the helical orbits in the Tevatron*, in proceedings of *Particle Accelerator Conference (PAC07)*, p. 3874, 25 – 29 June 2007, Albuquerque, New Mexico, USA.

[29] A. Valishev, et al, *Simulation of beam-beam effects and Tevatron experience*, in proceedings of *European Particle Accelerator Conference (EPAC08)*, 23 – 27 June 2008, Genoa, Italy.

[30] http://www-bd.fnal.gov/SDAViewersServlets/valishev_sa_catalog2.html

[31] A. Valishev and G. Stancari *Results of head-on beam-beam compensation studies at the Tevatron,* in proceedings of *Particle Accelerator Conference (PAC11)*, 28 March – 1 April 2011, New York, New York, USA.

[32] G. Stancari, G. Annala, V. Shiltsev, D. A. Still, A. Valishev, L. G. Vorobiev, *Experimental study of magnetically confined hollow electron beams in the Tevatron as collimators for intense high-energy hadron beams,* in proceedings of *Particle Accelerator Conference (PAC11)*, 28 March – 1 April 2011, New York, New York, USA.